\documentclass[prl,onecolumn,superscriptaddress,nofootinbib,12pt]{revtex4-2}
\usepackage{url}

\usepackage{float}
\usepackage{epsfig}
\usepackage{amsfonts}
\usepackage{graphicx}
\usepackage{amsmath}
\usepackage{amssymb}
\usepackage{color}
\usepackage{bbm}
\usepackage{dcolumn}% Align table columns on decimal point
\usepackage{bm}% bold math
\usepackage{ulem}
\usepackage{mathrsfs}
\usepackage{bbold}
\usepackage{datetime}
\usepackage{nameref}

\usepackage{rotating}
\usepackage[usenames,dvipsnames]{xcolor}
\usepackage[colorlinks=true,citecolor=Red,linkcolor=Green,urlcolor=Green]{hyperref}
\usepackage{lipsum} % for mock text
\usepackage{etoolbox} % for \appto
\usepackage[capitalize]{cleveref}
\usepackage{extarrows}

\def\bc{\begin{center}}
\def\nno{\nonumber}
\def\ec{\end{center}}
\def\be{\begin{eqnarray}}
\def\ee{\end{eqnarray}}
\definecolor{dyellow}{rgb}{1.,0.8,.0}
\definecolor{myblue}{rgb}{.1,.1,.7}
\definecolor{dcyan}{rgb}{.0,.6,.6}
\definecolor{dmagenta}{rgb}{0.6,0.0,0.6}
\definecolor{brown}{rgb}{0.6,0.2,0.}
\definecolor{darkblue}{rgb}{.0,.0,0.5}
\definecolor{darkred}{rgb}{0.75,0.0,0.0}
\definecolor{orange}{rgb}{1.,.6,.0}
\definecolor{dorange}{rgb}{0.8,.4,.0}
\definecolor{darkgreen}{rgb}{0.0,0.6,0.0}
\definecolor{purple}{rgb}{.4,.0,.4}
\definecolor{lightgrey}{rgb}{0.7, 0.7, 0.7}
\definecolor{grey}{rgb}{0.4, 0.4, 0.4}
\usepackage{geometry}
\usepackage[font=small,labelfont=bf,justification=raggedright]{caption}

\newcommand{\xdownarrow}[1]{%
  {\left\downarrow\vbox to #1{}\right.\kern-\nulldelimiterspace}
}
\newcommand{\xuparrow}[1]{%
  {\left\uparrow\vbox to #1{}\right.\kern-\nulldelimiterspace}
}

\definecolor{myred}{RGB}{189, 38, 49}

\begin{document}
\title{Measuring Renyi Entropy in Neural Network Quantum States}
\author{Han-Qing Shi} \email{by2030104@buaa.edu.cn}
\affiliation{Center for Gravitational Physics, Department of Space Science, Beihang University, Beijing 100191, China}
\author{Hai-Qing Zhang} \email{hqzhang@buaa.edu.cn}
\affiliation{Center for Gravitational Physics, Department of Space Science, Beihang University, Beijing 100191, China}
\affiliation{Peng Huanwu Collaborative Center for Research and Education, Beihang University, Beijing 100191, China}

\begin{abstract}
We compute the Renyi entropy in a one-dimensional transverse-field quantum Ising model in the ground state and in the state after a linear quench, by employing a swapping operator acting on the states which are prepared from the neural network methods. In the static ground state, Renyi entropy can uncover the critical point of the quantum phase transition from paramagnetic to ferromagnetic. At the critical point, the relation between the Renyi entropy and the subsystem size satisfies the predictions from conformal field theory. In the dynamical case, we find coherent oscillations of the Renyi entropy after the end of the linear quench. These oscillations have universal frequencies which may come from the superpositions of excited states. The asymptotic form of the Renyi entropy implies a new length scale away from the critical point. This length scale is also verified by the overlap of the reduced Renyi entropy against the dimensionless subsystem size. 
\end{abstract}

\maketitle

\section{Introduction}
Entanglement entropy (i.e., von Neumann entropy, defined as $S_1$ below) is one of the fascinating aspects in quantum physics, stimulating the developments of the quantum information and quantum computation \cite{nielsen2000quantum}. Measuring entanglement entropy can reveal some surprising correlations between far-way separated systems, making it as the most fascinating and mysterious physical quantity in quantum physics. Therefore, to study the properties of the entanglement entropy in quantum many-body systems or quantum field theory becomes an intriguing task \cite{Calabrese:2004eu,Calabrese:2009qy}. Renyi entropy ($S_n$, defined below), a generalized version of the entanglement entropy, has attracted much attention recently \cite{principe2010information}, since Renyi entropy can encode much more information than entanglement entropy in studying the entanglement spectrum of the density matrix \cite{li2008entanglement}. In the high energy communities, entanglement entropy and Renyi entropy have also attracted much attention in recent years due to the holographic duality  \cite{maldacena1999large}, from which one can use the Ryu-Takayanagi formula to calculate them \cite{ryu2006holographic,Hung:2011nu}. 

Recently, neural networks with machine learning methods has become a powerful tool in solving quantum many-body problems \cite{Carrasquilla:2017aa,carleo2019machine}, ranging from static ground states to time-dependent evolutions. Besides these successes in quantum physics, it is also been adopted in a broad spectrum of physics, from computational physics \cite{cuomo2022scientific} to high energy physics \cite{d2019learning} and cosmology \cite{hezaveh2017fast}. The wide range of its applications is due to the high efficient representing power of neural networks, which can characterize the states of the system with acceptable number of parameters \cite{iten2020discovering}. Therefore, neural networks makes the overwhelming complexity computationally tractable. Computation of entanglement in a quantum system is formidable because of its exponential complexity of states in the Hilbert space. Therefore, to study the entanglement of a quantum system by virtue of the neural networks becomes an intriguing task.

In this paper, we will focus on studying the second Renyi entropy, i.e. $S_2$, of a one-dimensional transverse-field quantum Ising model (TFQIM) by utilizing the neural network methods. We prepare the neural-network quantum states (NQS) of TFQIM with the machine learning methods \cite{carleo2017solving}. Then we transform the studying of $S_2$ by computing the expectation value of a swapping operator \cite{hastings2010measuring}. First, we study the Renyi entropy in static ground states. By varying the transverse magnetic field, the peak of the Renyi entropy appears at the critical value of the magnetic field, which reflects a quantum phase transition from paramagnetic to ferromagnetic. Therefore, Renyi entropy can serve as an order parameter to disclose the phase transition. As the transverse magnetic field goes to zero, Renyi entropy tends to $\ln2$ which is consistent with the two-degenerated vacuum states in the pure ferromagnetic phase, i.e., the spins all point up or down \cite{Calabrese:2004eu}. We also study the relation between the Renyi entropy and the subsystem size at the critical point, and find that it satisfies the theoretical predictions in conformal field theory \cite{Calabrese:2004eu}. The consistency of the studying of $S_2$ in the ground states reflects that the neural network methods apply well to the quantum phase transition of TFQIM at least in static case.

Later, we extend the neural network methods to the dynamical case. We study the Renyi entropy in a dynamical phase transition induced by a linear quench of the transverse magnetic field. This is a paradigmatic model in studying the formation of one-dimensional defects (kinks) from the celebrated Kibble-Zurek mechanism (KZM) \cite{kibble1976topology,zurek1985cosmological}. From KZM, there is an adiabatic-impulse-adiabatic process as the magnetic field traverses from paramagnetic to ferromagnetic state. In the impulse regime nearby the critical point, the system is excited rather than following the original ground state. Therefore, Renyi entropy in this case is assumed to be different from those in the static case. Specifically, we quench the transverse magnetic field from greater than critical values to zero, and then let the system evolve freely. We find that superpositions of different symmetry-broken states result in the coherent oscillations of the Renyi entropy after the end of the quench. The periods (or frequencies) of these oscillations for different quench rates are identical, which implies the degeneracy of energy levels as magnetic field vanishes. Moreover, we find that there emerges a new length scale $\xi$ in this free evolution regime. This length scale is different from the one in the impulse regime, i.e., $\hat\xi$, which is the correlation length used in KZM to predict the number density of the defects. We find that the relation between the asymptotic behavior of the Renyi entropy and the quench rate satisfies the theoretical formula if we recognize $\xi$ as the new length scale. Besides, the collapsing of relations between the reduced Renyi entropy and the reduced dimensionless subsystem size also supports the above asserts that in the free evolution regime $\xi$ is the new length scale.  

\section{Basic set-up}
\subsection{Transverse-field quantum Ising model}
The Hamiltonian of one-dimensional TFQIM with $N$ sites is \cite{franchini2017introduction}
\begin{equation}
	H=-J\sum^{N}_{i=1}(\sigma_i^z\sigma_{i+1}^z+h\sigma_i^x)
\end{equation}
where $\sigma^{z}_i$ and $\sigma^x_i$ are the Pauli matrices at the site $i$, $J$ represents the coupling strengths between the nearest-neighbor sites while $h$ denotes the strength of the transverse magnetic field. We adopt the periodic boundary conditions (PBC), i.e., $\vec\sigma_{N+1}=\vec\sigma_1$. In this paper we set $J=1$. 
In the ground state, there is a quantum phase transition at the critical value $h_c=1$. (We will only consider $h\geq0$ since $h\leq0$ has similar phenomenon.)  As $h\gg1$, the system is in a paramagnetic phase with spins all pointing along $x$-direction. On the contrary, as $h<1$, the system is in a ferromagnetic phase. In particular, when $h=0$ the system falls into a complete ferromagnetic phase in which the system is in a two-degenerated state with all spins either pointing up or down along the $z$-direction. 

This phase transition can also be captured by the Renyi entropy \cite{Calabrese:2004eu}. If the system is taken in a pure state $|\Psi\rangle$, then one can separate it into two subsystems $A$ and $B$. The reduced density matrix of $A$ system is defined as $\rho_A={\rm tr}_B|\Psi\rangle\langle\Psi|$, then the Renyi entropy is defined as   
\begin{eqnarray}
	%S_A(\rho_A)=&-{\rm Tr}_A(\rho_A\ln\rho_A)\\
	S_n(\rho_A)=&\frac{1}{1-n}\ln\left[{\rm Tr}(\rho_A^n)\right]
\end{eqnarray}
where $S_n(\rho_A)$ is the $n$-th Renyi entropy of the subsystem $A$. As $n\to1$, it reduces to the von Neumann entropy $S_1=-{{\rm Tr}(\rho_A\ln\rho_A)}$. For an infinitely large system, Renyi entropy will diverge at the critical point $h_c=1$, while for a finite system, it has a finite peak at this point. As $h=0$, due to the $Z_2$ symmetry of the complete ferromagnetic system, the value of Renyi entropy should be $\ln 2$. As $h>1$, the system is in a paramagnetic phase and Renyi entropy will decrease and then vanish in the limit of $h\to\infty$.
 
However, for a time-dependent $h(t)$, the studying of Renyi entropy in the dynamical phase transition of TFQIM is rare. We assume that the system is undergoing a linear quench of the transverse magnetic field from paramagnetic to ferromagnetic phase, i.e.,
\begin{equation}\label{quench}
	h(t)=-\frac{t}{\tau_Q},~~~t\in [-T,0].
\end{equation}
where $\tau_Q$ is the quench rate and $-T$ is the initial time of the quench. At the initial time, $h\gg1$ and the system is prepared in the paramagnetic phase; By quenching the system across the critical point $h_c=1$, dynamical symmetry-breaking will bring out the topological defects (kinks) with the density depending on the quench rate $\tau_Q$ according to the KZM \cite{kibble1976topology,zurek1985cosmological}.  This quantum phase transition inevitably excites the system and the final state is a superposition of the excited states with different symmetry-broken domains \cite{dziarmaga2005dynamics,dziarmaga2022coherent}. Therefore, we can readily imagine that Renyi entropy in the dynamical case will be different from that in static case. 

\subsection{Machine learning with neural network states}

We utilize the restricted Boltzmann machines (RBM) as the neural networks to represent the states of the TFQIM \cite{carleo2017solving}. This network has two layers, one includes $N$ visible neurons $s_j$ (i.e., the spins at site $j$) and the other has $M$ hidden neurons $h_i$. The visible neurons are assigned as input with the directions of spins as $s(\uparrow)=1$ and $s(\downarrow)=-1$. The hidden neurons can be assigned the same values $h_i=\{1,-1\}$. According to the structure of RBM, we can write down the expression of NQS with the neural network parameters $\mathcal{W}=\{a,b,w\}$ \cite{carleo2017solving},
\begin{equation}
	\Psi_{\rm NQS}(s,\mathcal{W})=\sum_{\{h_i\}}e^{\left.\sum_j a_j s_j + \sum_i b_i h_i+ \sum_{ij}w_{ij}h_i s_j\right.}.
\end{equation}

The dynamics we considered is an evolution from an initial ground state with strong transverse magnetic field to a state with vanishing transverse magnetic field. In order to obtain this initial ground state, we need to train a beginning wave function whose parameters are random complex numbers from machine learning algorithms, which is realized through minimizing the expectation value of the energy with the Stochastic Reconfiguration (SR) method \cite{sorella2001generalized}.
After preparing the initial ground state, the NQS need to satisfy the time-dependent Schr\"odinger equation according to the linear quench \eqref{quench}. To this end, we utilize the time-dependent variational principle (TDVP) to solve the time-dependent neural network parameters $\mathcal{W}(t)$. In each iteration, we need to minimize the Fubini-Study distance 
\begin{equation}
	\mathcal{D}(\mathcal{W})=\arccos\left[\frac{\langle \partial_t \Psi|H\Psi \rangle \langle H\Psi|\partial_t \Psi \rangle}{\langle \partial_t \Psi|\partial_t \Psi \rangle \langle H\Psi|H\Psi \rangle} \right]^{1/2}
\end{equation}
which describes the difference between the exact time evolution and variational evolution.
Eventually, it can yield an evolution equation about the parameters $\mathcal{W}$ \cite{carleo2017solving},
\begin{equation}
	S_{k'k}\dot{\mathcal{W}}_k=-iF_{k'}
\end{equation} 
where $\dot\ $ represents time derivative, $S_{k'k}=\langle O^*_{k'}O_k \rangle-\langle O^*_{k'}\rangle\langle O_k\rangle$ is a covariance matrix of the operator $O_k=\tfrac{\partial\ln\Psi}{\partial \mathcal{W}_k}$ and $F_{k'}=\langle O^*_{k'}E_{\rm loc} \rangle-\langle O^*_{k'}\rangle\langle E_{\rm loc}\rangle$ is the generalized forces with $E_{\rm loc}=\tfrac{\langle s|H|\Psi\rangle}{\Psi(s)}$. According to TDVP, the updated parameter is $\mathcal{W}(t+\delta t)=\mathcal{W}(t)+\dot{\mathcal{W}}\delta t$ in each iteration.

\subsection{Renyi entropy from swapping operation}
 For simplicity, we call $S_2$ as Renyi entropy in the following. We will utilize a swapping operation $S_{\rm wap}$ to measure $S_2$ from NQS \cite{hastings2010measuring,torlai2018neural}. 
Assume that one can construct the state $|\Psi\rangle$ of the system from two parts' basis $|\alpha\rangle$ and $|\beta\rangle$, where $|\alpha\rangle$ is a complete basis of states in the part $A$, and $|\beta\rangle$ is a complete basis of state in the complement part $B$. Therefore, the whole state of the system can be decomposed in the product of basis as $|\Psi\rangle=\sum_{\alpha\beta}C_{\alpha\beta}|\alpha\rangle|\beta\rangle$. The swapping operator $S_{\rm wap}$ will exchange the states in the part $A$ between the original and the copied one, such that,
\begin{eqnarray}
	&&S_{\rm wap}\left(\sum_{\alpha_1 \beta_1}C_{\alpha_1\beta_1}|\alpha_1\rangle|\beta_1\rangle\right)\otimes\left(\sum_{\alpha_2 \beta_2}D_{\alpha_2\beta_2}|\alpha_2\rangle|\beta_2\rangle\right)\nonumber\\
	&&=\sum_{\alpha \beta}C_{\alpha_1\beta_1}D_{\alpha_2\beta_2}|\alpha_2\rangle|\beta_1\rangle\otimes|\alpha_1\rangle|\beta_2\rangle
\end{eqnarray}
According to this definition, one can link the expectation value of $\langle S_{\rm wap}\rangle$ to the Renyi entropy,
\begin{eqnarray}
	&&\langle\Psi\otimes\Psi|S_{\rm wap}|\Psi\otimes\Psi\rangle=\sum_{\alpha\beta}C_{\alpha_1\beta_1}{C}^*_{\alpha_2\beta_1}C_{\alpha_2\beta_2}{C}^*_{\alpha_1\beta_2}\nonumber\\
	&&=\sum_{\alpha_1\alpha_2}[\rho_A]_{\alpha_1\alpha_2}[\rho_A]_{\alpha_2\alpha_1}={\rm Tr}(\rho_A^2)=e^{-S_2}
\end{eqnarray}
where $[\rho_A]_{\alpha_1\alpha_2}=\sum_{\beta_1}C_{\alpha_1\beta_1}{C}^*_{\alpha_2\beta_1}$ is the element of the reduced density matrix $\rho_A$. The coefficient $C$ and ${C}^*$ can be obtained by samplings from the NQS with quantum Monte-Carlo methods, in our paper we use $1000$ samplings.

\section{Results}
\subsection{Renyi entropy in static ground state}
\begin{figure}[h]
	\centering
		\includegraphics[trim=0.3cm 4.35cm 1.9cm 6.cm, clip=true, scale=0.38]{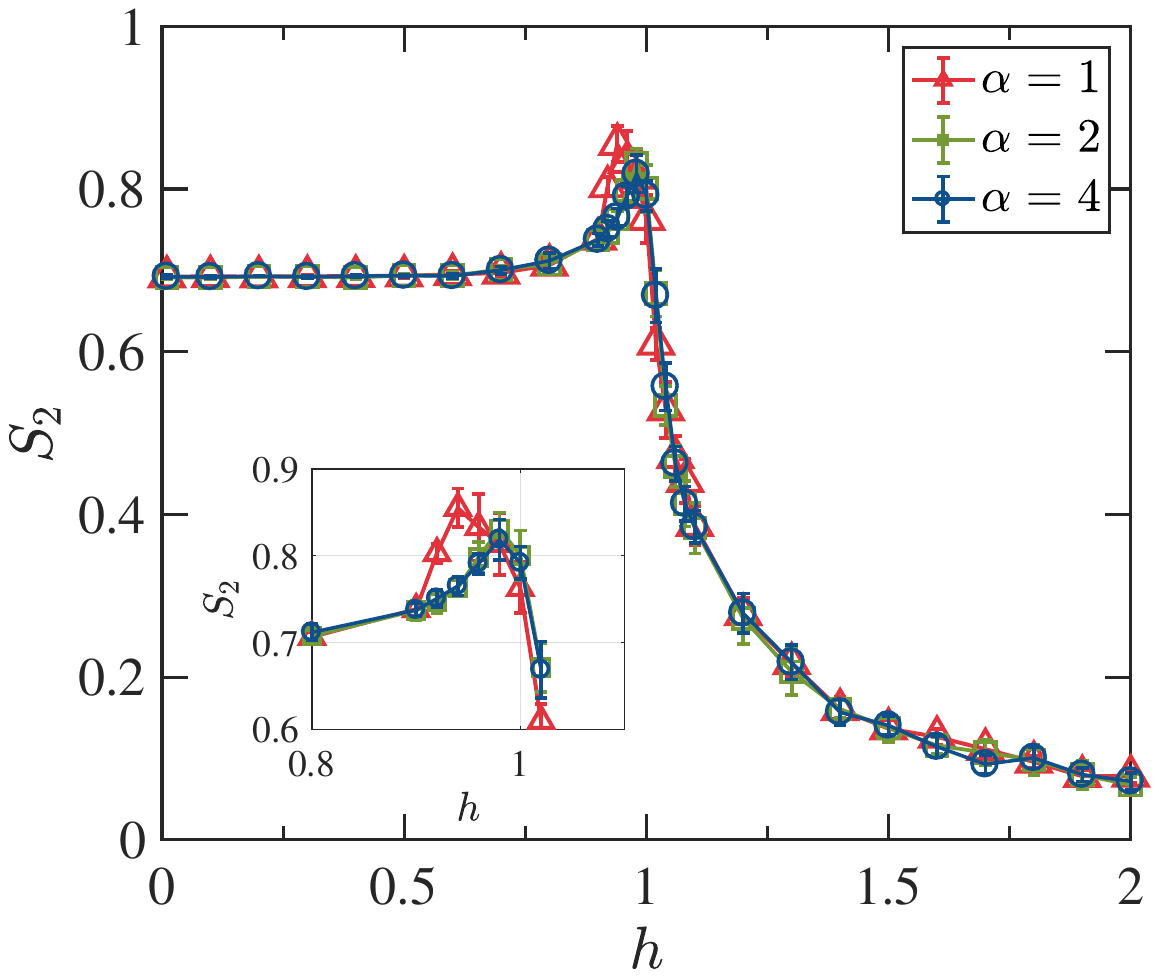}
		\put(-210,196){\footnotesize{(a)}}
		~\includegraphics[trim=0.7cm 6.5cm 2.6cm 7.3cm, clip=true, scale=0.47]{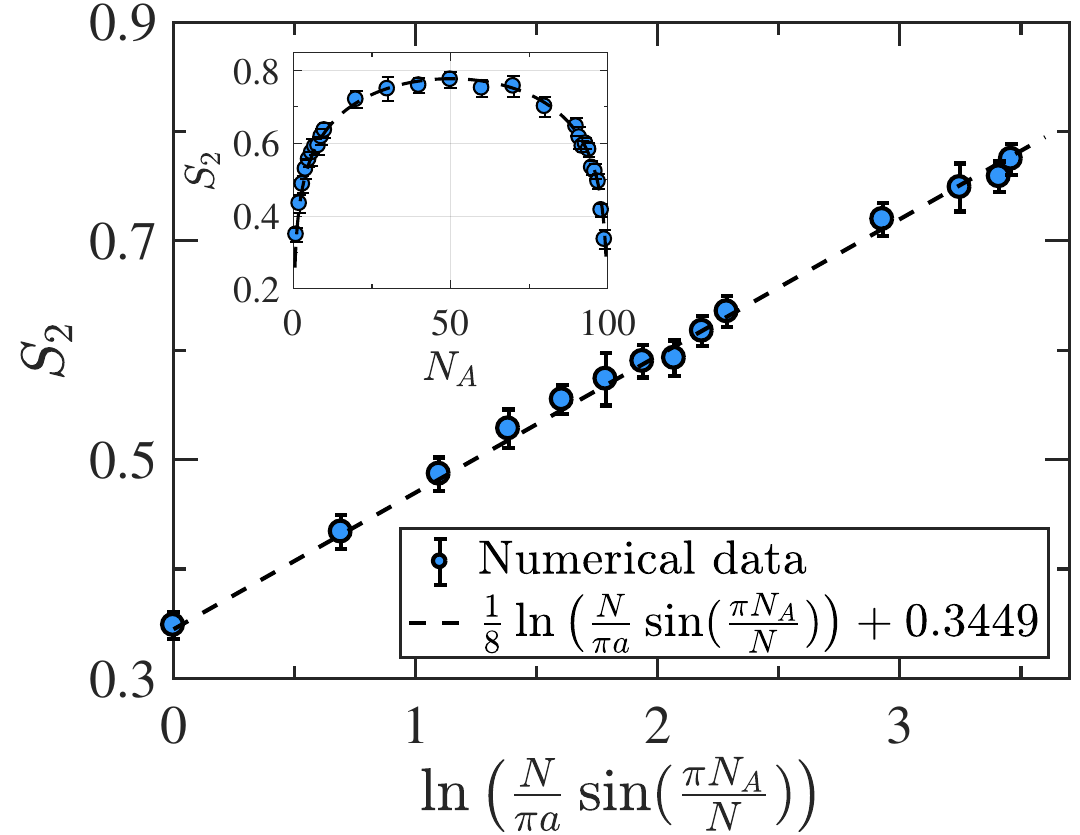}
		\put(-240,196){\footnotesize{(b)}}
		\caption{(a) Renyi entropy for half-size subsystem in static ground states against the transverse-field strength with network size $\alpha=1,2,4$. Inset plot shows that the peak of Renyi entropy will be closer to the critical point $h_c=1$ as increasing the size $\alpha$; (b) Linear relation between the Renyi entropy and $\ln\left(\frac{N}{\pi a}\sin\left(\frac{\pi N_A}{N}\right)\right)$ for $N_A\leq50$ at the critical point, satisfying the theoretical results \eqref{naln}. The inset plot shows the relation between $S_2$ and the subsystem size $N_A$ ($N_A\leq100$) in the ordinary coordinates. The error bars stand for the standard errors, and the dashed lines are the fitting lines of the numerical data.}
		\label{static}
\end{figure}

First, we study the Renyi entropy from the neural networks for the static one dimensional TFQIM with $N=100$ sites. In Fig.\ref{static}(a), we show the Renyi entropy against different transverse magnetic field strength $h$ for various network size $\alpha$. We define $\alpha$ as the ratio between the hidden and visible neurons $\alpha=M/N$. Therefore, bigger $\alpha$ represents higher accuracy of the neural networks. Theoretically, the entropy will diverge at the critical point $h_c=1$ in the infinite-size system. While in the finite system, a finite peak will replace the divergence. Here, the Renyi entropy is for the half-size subsystem, i.e., an interval with $50$ sites. We indeed see a peak around $h_c=1$. When increasing $\alpha$, the peak will move closer to the critical point position $h_c=1$, see the inset plot in Fig.\ref{static}(a). As $h\to0$, the Renyi entropy goes to the theoretical predicted value $\ln2\approx0.6931$ due to the expectation that there are two possible ground states with opposite directions of the magnetization. While $h\to\infty$, it decreases and tends to zero due to the expectation that there is only one configuration of the magnetization all pointing to $x$-direction. 

At the critical point, the dependence of Renyi entropy $S_n$ to the size $N_A$ of the subsystem $A$ is (with PBC)
\cite{Calabrese:2009qy}
\be\label{naln}
S_n(N_A)=\frac{c}{6}\left(1+\frac 1n\right)\ln\left(\frac{N}{\pi a}\sin\left(\frac{\pi N_A}{N}\right)\right)+{\rm const.}
\ee
where $c=1/2$ is the central charge in TFQIM, while $a$ is the lattice spacing and was set to $a=1$ throughout this paper. Fig.\ref{static}(b) exhibits a linear relation between $S_2$ and the logarithmic term for $N_A\leq50$. The slope $1/8$ is consistent with the theoretical prediction \eqref{naln}. In the inset plot we show the regular relation between $S_2$ and the size $N_A$ in the whole range. The numerical data are symmetric under $N_A\leftrightarrow 100-N_A$, and the maximum value appears at $N_A=50$. The results of Fig.\ref{static} are consistent with theoretical predictions \cite{Calabrese:2009qy}, indicating that the machine learning methods are working well for the static ground state in TFQIM \cite{shi2019neural,shi2022learning}.

%\onecolumngrid\
%\twocolumngrid\

\begin{figure}[h]
	\begin{center}
		\includegraphics[trim=2.4cm 0.5cm 3cm 1cm, clip=true, scale=0.5]{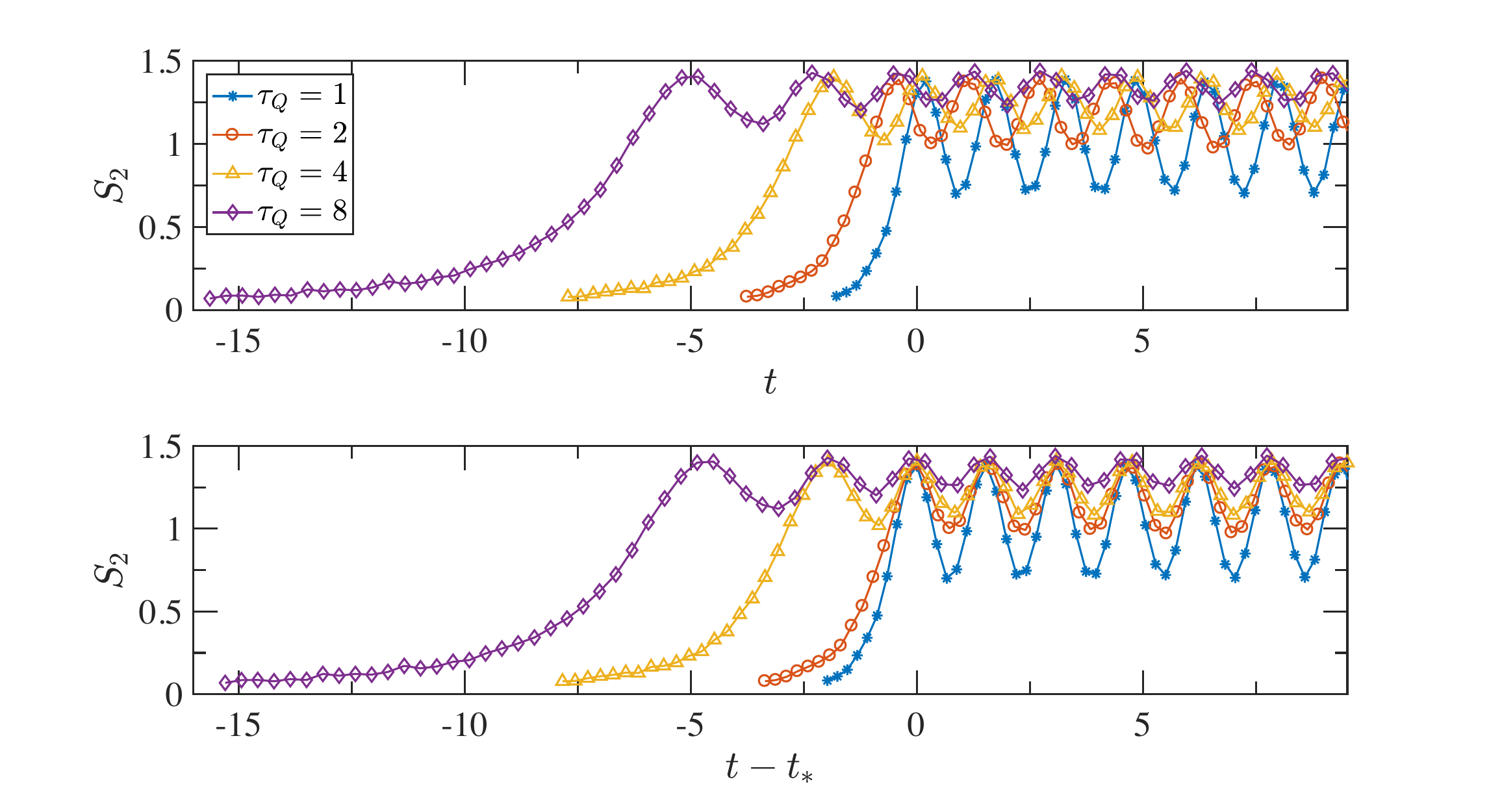}
		\put(-488,270){\footnotesize{(a)}}
		\put(-488,130){\footnotesize{(b)}}
		\caption{(a) Time evolution of Renyi entropy for the various quench rates $\tau_Q=1, 2, 4$ and $8$. As $t<0$, the system undergoes a linear quench from $h=2$ to $h=0$; After $t=0$, we keep the system to evolve with zero transverse magnetic field for a while. Renyi entropy begin to oscillate after the ramp; (b) In order to compare the periods of the oscillations for different quench rates, we shift the Renyi entropy by the phase differences $t_*$. Then we see that the periods for these oscillations are almost the same, which is close to $\pi/2$. }
		\label{s2_nt}
	\end{center}
\end{figure}

\subsection{Renyi entropy in quenched dynamics}
In the time-dependent evolution, we consider a linear quench of the transverse magnetic field as in Eq.\eqref{quench}. 
At the initial time we set $h=2$, where the system is prepared in the paramagnetic ground state. In the numerics, time step is set to $\Delta t=10^{-3}$. Then we quench the magnetic field linearly to zero, and keep the system to evolve freely with $h=0$ for a while. During the quench, the system will be excited to a ferromagnetic state with kink configurations due to the KZM. The superpositions of the excited states will result in different behaviors of the Renyi entropy compared to the static case. 

We show the time evolution of $S_2$ in Fig.\ref{s2_nt}(a) for various quench rates. $t=0$ is the instant for ending the quench. We see that Renyi entropy will grow in time and then enter an oscillatory phase. For faster quench (smaller $\tau_Q$), the ramp is steeper. However, in the oscillatory phase the periods of the oscillations are almost the same for different quench rates. In the Fig.\ref{s2_nt}(b) we slightly shift the Renyi entropy by a phase difference $t_*$ in order to compare their periods conveniently. Specifically, $t_*(\tau_Q=1)=0.2$, $t_*(\tau_Q=2)=-0.4$, $t_*(\tau_Q=4)=0.12$, and $t_*(\tau_Q=8)=-0.34$. We can clearly see that their periods are almost similar and are roughly $T_{\rm period}\approx1.5867$. Interestingly, we find that this period is coincident with that of the coherent oscillatory behavior of transverse magnetization in \cite{dziarmaga2022coherent}, in which the analytical period was $\pi/2$ which is very close to our numerical results. The coherent oscillations are due to the superpositions of the different symmetry-broken excited states. Therefore, we speculate that the oscillatory behavior of $S_2$ in our paper may come from the superpositions of the excited states after the end of the quench.  

\begin{figure}[h]
	\begin{center}
		\includegraphics[trim=1.3cm 6.5cm 2cm 7.5cm, clip=true, scale=0.5]{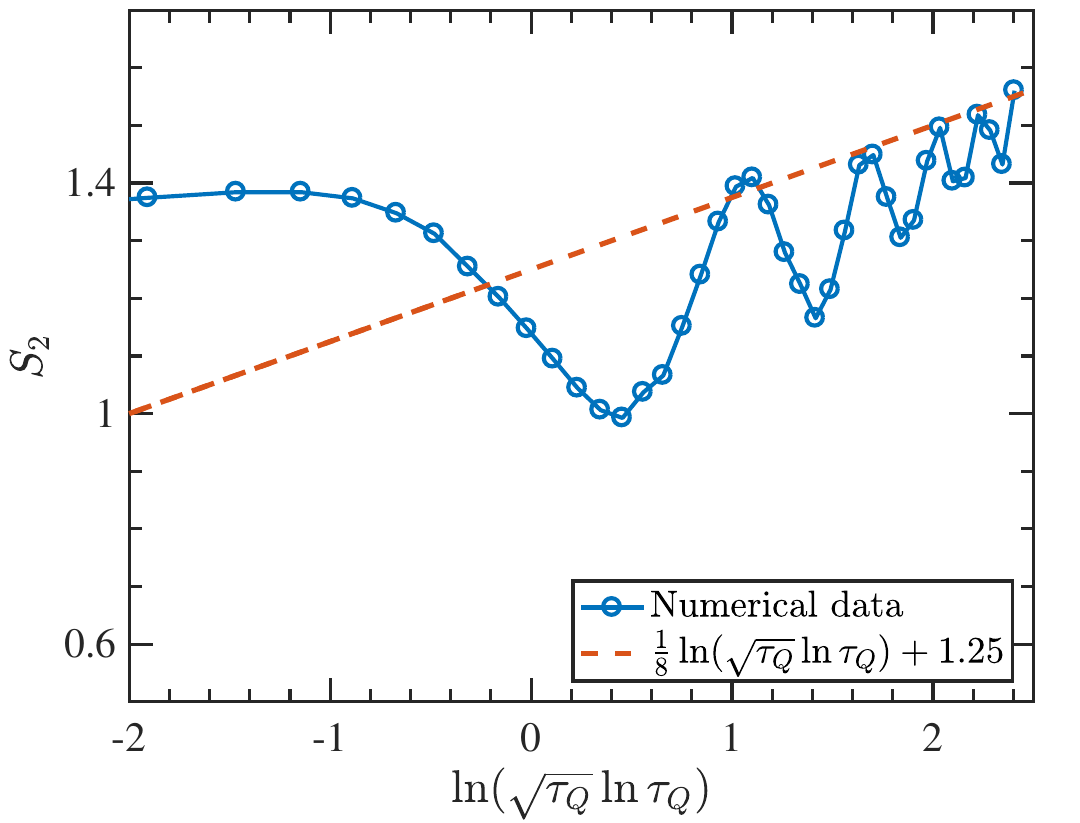}
		\caption{Renyi entropy at the end of quench with respect to the logarithmic function of quench rate $\ln(\sqrt{\tau_Q}\ln\tau_Q)$. The asymptotic form of $S_2$ as $\tau_Q\to\infty$ is fitted by the red dashed line, which is consistent with the Eq.\eqref{s2tq}. }
		\label{s2_tq}
	\end{center}
\end{figure}

At the end of the quench, i.e., at $t=0$, we can read off the relation between the Renyi entropy and the logarithmic function of quench rate, which is shown in Fig.\ref{s2_tq}. We already see from Fig.\ref{s2_nt}(a) that the Renyi entropy is oscillatory at $t=0$, therefore, it is easy to deduce that $S_2$ at the end of quench is not monotonic against the quench rate. From Fig.\ref{s2_tq} we indeed see that $S_2$ is oscillatory as well with respect to $\tau_Q$. However, as $\tau_Q$ grows, this oscillation becomes smaller, which is reflected in Fig.\ref{s2_nt}(a) that the amplitudes of $S_2$ become smaller as $\tau_Q$ is greater. \footnote{In numerics, the largest quench rate we used is $\tau_Q=16$ in order to balance the precisions and time consumptions of the codes. } The asymptotic form of the Renyi entropy as $\tau_Q\to\infty$ can be fitted by the theoretical formula in static case. At the instant $t=0$ the system is away from the critical point. According to \cite{cincio2007entropy}, there emerged a new length scale $\xi\simeq\sqrt{\tau_Q}\ln\tau_Q$ in the end of the quench.  Therefore, the Renyi entropy for an interval can be predicted from the length scale $\xi$ as \cite{Calabrese:2009qy},
\be\label{s2tq}
S_n&=&2\times\frac{c}{12}\left(1+\frac1n\right)\ln\left(\frac{\xi}{a}\right)+{\rm const.}\nno\\
\Rightarrow S_2&=&\frac18\ln(\sqrt{\tau_Q}\ln\tau_Q)+{\rm const.}
\ee
Note that multiplying a prefactor $2$ is because the interval has two boundary points \cite{Calabrese:2009qy}. The asymptotic behavior of $S_2$ is denoted as the red dashed line in Fig.\ref{s2_tq}, which is consistent with the theoretical formula \eqref{s2tq}.

\begin{figure}[t]
	\begin{center}
		\includegraphics[trim=1.cm 6.6cm 2.cm 7.2cm, clip=true, scale=0.44]{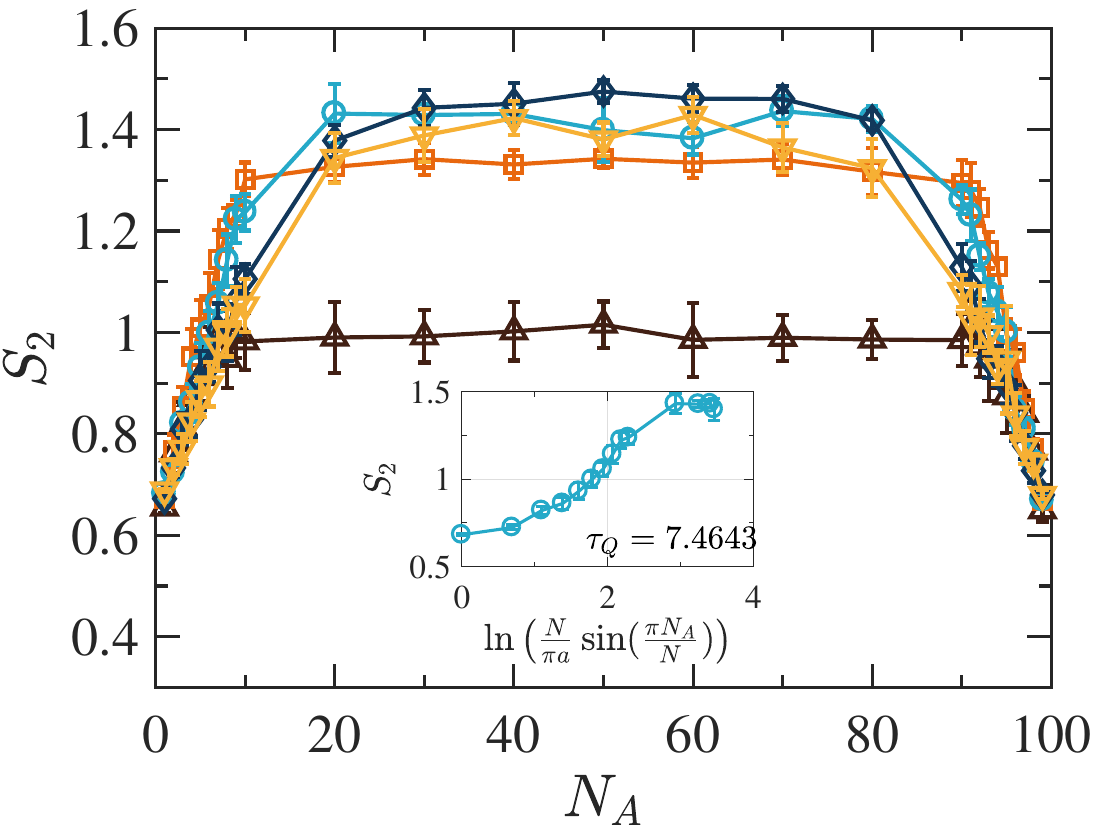}
		\put(-230,176){\footnotesize{(a)}}
		~\includegraphics[trim=1.cm 6.6cm 2.cm 7.2cm, clip=true, scale=0.44]{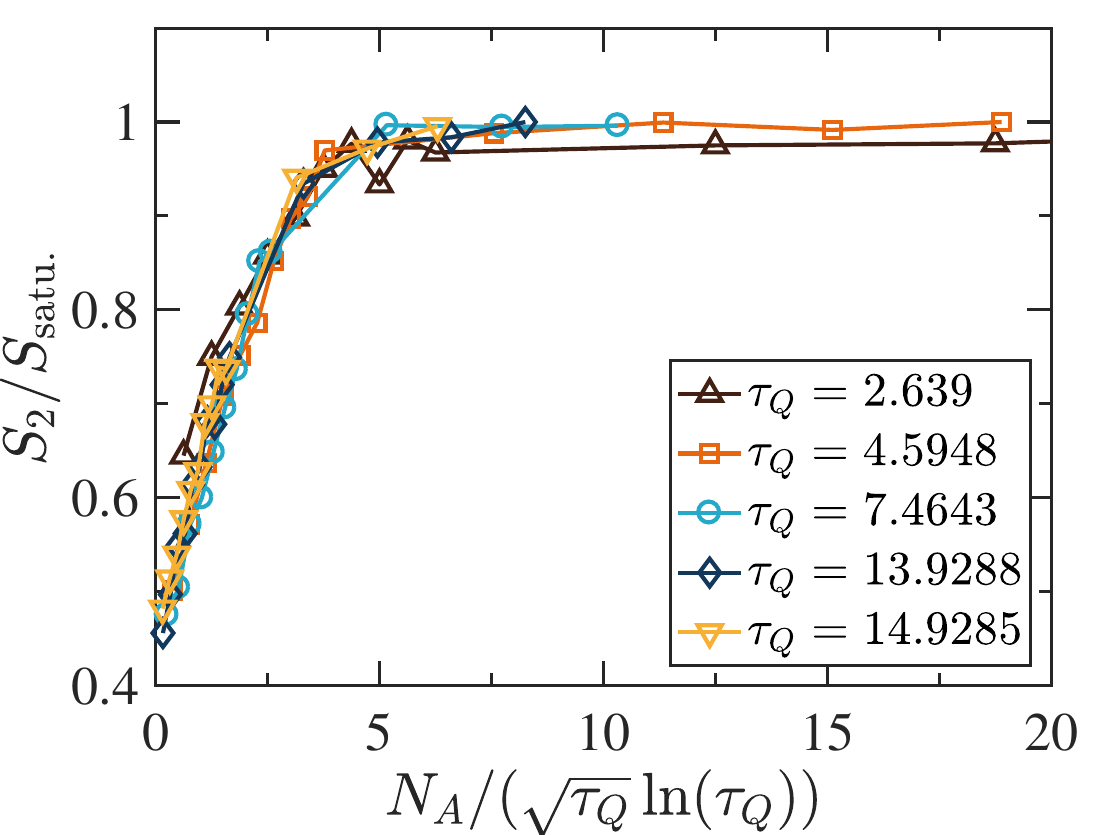}
		\put(-230,176){\footnotesize{(b)}}
		\caption{(a) Renyi entropy against the size of the interval $N_A$ for several quench rates. Renyi entropy will increase with $N_A$ and reach saturation quickly before the half of system. The error bars represent the standard errors. The inset is an example for $\tau_Q=7.4643$ to show that the relation Eq.\eqref{naln} in static case does not work in the quenched case; (b) The relation between the reduced Renyi entropy and the reduced length scale. They collapse together for each quench rate.  }
		\label{s2_na}
	\end{center}
\end{figure}

In the Fig.\ref{s2_na}(a) we show the relation of the Renyi entropy with respect to the size of the subsystem at the end of quench $t=0$, for various quench rates. It is obvious that this relation should be different from the static case, i.e., Eq.\eqref{naln}. As an example, we show in the inset plot the relation between $S_2$ and $\ln\left(\frac{N}{\pi a}\sin\left(\frac{\pi N_A}{N}\right)\right)$ for $\tau_Q=7.4643$. It is away from the linear relation Eq.\eqref{naln}. From Fig.\ref{s2_na}(a) we can still see the symmetry under $N_A\leftrightarrow100-N_A$. Besides, in the range $N_A\in[0,50]$, Renyi entropy grows as the subsystem size grows, and then saturates nearby $N_A=50$. We define the Renyi entropy at saturation as $S_{\rm satu.}$. 
As was mentioned above, at this case the length scale is $\xi\simeq\sqrt{\tau_Q}\ln\tau_Q$. Therefore,  the ratio $N_A/(\sqrt{\tau_Q}\ln\tau_Q)$ is a reduced dimensionless scale. We again define a reduced Renyi entropy which is the ratio between $S_2$ and $S_{\rm satu.}$ for each quench rate.  Their relations can be found in Fig.\ref{s2_na}(b). We see that for each quench rate they overlap very well, which implies that the length scale $\xi\simeq\sqrt{\tau_Q}\ln\tau_Q$ at $t=0$ is reasonable.

\section{Conclusions and Discussions}
 We studied the second Renyi entropy $S_2$ for the one-dimensional TFQIM with neural networks in both static ground state and quenched dynamics. By adopting the swapping operation, we transform the computation of $S_2$ to calculating the expectation value of the swapping operator $S_{\rm wap}$. In the static case, the peak of the Renyi entropy can reveal the critical point of the quantum phase transition from paramagnetic to ferromagnetic. Therefore, Renyi entropy in this sense can play a role of order parameter for uncovering phase transitions. The relation between Renyi entropy and the size of the subsystem matches the theoretical predictions very well, which verifies the accuracy of the neural network methods. 

In the quenched dynamics, we linearly quenched the transverse magnetic field from $h>1$ to $h=0$ and then let the system evolve freely. We found the coherent oscillations of the Renyi entropy in the free evolutions. The oscillation periods are identical to $\pi/2$ which was consistent with the coherent oscillations of the transverse magnetization studied in existing literatures. We interpreted this oscillation behavior as the superpositions of the different symmetry-broken excited states. At the end of the quench, the system embodies a new length scale as $\xi\simeq\sqrt{\tau_Q}\ln\tau_Q$, different from the scale $\hat\xi\simeq\sqrt{\tau_Q}$ in the impulse regime which is nearby the critical point. We verified this new length scale from the asymptotic behavior of the Renyi entropy in the limit of large quench rate. Besides, the relation between the reduced Renyi entropy $S_2/S_{\rm satu.}$ and the reduced size of the subsystem $N_A/\xi$ collapse together, which indicates that the new length scale $\xi\simeq\sqrt{\tau_Q}\ln\tau_Q$ at the end of the quench is correct.

\section*{Acknowledgements}
We appreciate the helpful discussions with Marek Rams. This work was partially supported by the National Natural Science Foundation of China (Grants No.12175008).

\appendix
\setcounter{equation}{0}
\setcounter{figure}{0}
\setcounter{table}{0}
\setcounter{section}{0}
%%\setcounter{page}{1}
%%\makeatletter
\renewcommand{\theequation}{S\arabic{equation}}
\renewcommand{\thefigure}{S\arabic{figure}}
%%\renewcommand{\bibnumfmt}[1]{[S#1]}
%%\renewcommand{\citenumfont}[1]{S#1}

%\section{Appdendix I: Numerical Methods}
%\label{appa}

\normalem
\bibliographystyle{ieeetr}
\bibliography{ref1.bib}

\begin{thebibliography}{10}

\bibitem{nielsen2000quantum}
M.~A. Nielsen and I.~L. Chuang, ``Quantum information and quantum
  computation.'' Cambridge University Press, 2000.

\bibitem{Calabrese:2004eu}
P.~Calabrese and J.~L. Cardy, ``{Entanglement entropy and quantum field
  theory},'' {\em J. Stat. Mech.}, vol.~0406, p.~P06002, 2004.

\bibitem{Calabrese:2009qy}
P.~Calabrese and J.~Cardy, ``{Entanglement entropy and conformal field
  theory},'' {\em J. Phys. A}, vol.~42, p.~504005, 2009.

\bibitem{principe2010information}
J.~C. Principe, {\em Information theoretic learning: Renyi's entropy and kernel
  perspectives}.
\newblock Springer Science \& Business Media, 2010.

\bibitem{li2008entanglement}
H.~Li and F.~D.~M. Haldane, ``Entanglement spectrum as a generalization of
  entanglement entropy: Identification of topological order in non-abelian
  fractional quantum hall effect states,'' {\em Physical review letters},
  vol.~101, no.~1, p.~010504, 2008.

\bibitem{maldacena1999large}
J.~Maldacena, ``The large-n limit of superconformal field theories and
  supergravity,'' {\em International journal of theoretical physics}, vol.~38,
  no.~4, pp.~1113--1133, 1999.

\bibitem{ryu2006holographic}
S.~Ryu and T.~Takayanagi, ``Holographic derivation of entanglement entropy from
  the anti--de sitter space/conformal field theory correspondence,'' {\em
  Physical review letters}, vol.~96, no.~18, p.~181602, 2006.

\bibitem{Hung:2011nu}
L.-Y. Hung, R.~C. Myers, M.~Smolkin, and A.~Yale, ``{Holographic Calculations
  of Renyi Entropy},'' {\em JHEP}, vol.~12, p.~047, 2011.

\bibitem{Carrasquilla:2017aa}
J.~Carrasquilla and R.~G. Melko, ``Machine learning phases of matter,'' {\em
  Nature Physics}, vol.~13, no.~5, pp.~431--434, 2017.

\bibitem{carleo2019machine}
G.~Carleo, I.~Cirac, K.~Cranmer, L.~Daudet, M.~Schuld, N.~Tishby,
  L.~Vogt-Maranto, and L.~Zdeborov{\'a}, ``Machine learning and the physical
  sciences,'' {\em Reviews of Modern Physics}, vol.~91, no.~4, p.~045002, 2019.

\bibitem{cuomo2022scientific}
S.~Cuomo, V.~S. Di~Cola, F.~Giampaolo, G.~Rozza, M.~Raissi, and F.~Piccialli,
  ``Scientific machine learning through physics--informed neural networks:
  Where we are and what's next,'' {\em Journal of Scientific Computing},
  vol.~92, no.~3, p.~88, 2022.

\bibitem{d2019learning}
R.~T. D'agnolo and A.~Wulzer, ``Learning new physics from a machine,'' {\em
  Physical Review D}, vol.~99, no.~1, p.~015014, 2019.

\bibitem{hezaveh2017fast}
Y.~D. Hezaveh, L.~P. Levasseur, and P.~J. Marshall, ``Fast automated analysis
  of strong gravitational lenses with convolutional neural networks,'' {\em
  Nature}, vol.~548, no.~7669, pp.~555--557, 2017.

\bibitem{iten2020discovering}
R.~Iten, T.~Metger, H.~Wilming, L.~Del~Rio, and R.~Renner, ``Discovering
  physical concepts with neural networks,'' {\em Physical review letters},
  vol.~124, no.~1, p.~010508, 2020.

\bibitem{carleo2017solving}
G.~Carleo and M.~Troyer, ``Solving the quantum many-body problem with
  artificial neural networks,'' {\em Science}, vol.~355, no.~6325,
  pp.~602--606, 2017.

\bibitem{hastings2010measuring}
M.~B. Hastings, I.~Gonz{\'a}lez, A.~B. Kallin, and R.~G. Melko, ``Measuring
  renyi entanglement entropy in quantum monte carlo simulations,'' {\em
  Physical review letters}, vol.~104, no.~15, p.~157201, 2010.

\bibitem{kibble1976topology}
T.~W. Kibble, ``Topology of cosmic domains and strings,'' {\em Journal of
  Physics A: Mathematical and General}, vol.~9, no.~8, p.~1387, 1976.

\bibitem{zurek1985cosmological}
W.~H. Zurek, ``Cosmological experiments in superfluid helium?,'' {\em Nature},
  vol.~317, no.~6037, pp.~505--508, 1985.

\bibitem{franchini2017introduction}
F.~Franchini {\em et~al.}, {\em An introduction to integrable techniques for
  one-dimensional quantum systems}, vol.~940.
\newblock Springer, 2017.

\bibitem{dziarmaga2005dynamics}
J.~Dziarmaga, ``Dynamics of a quantum phase transition: Exact solution of the
  quantum ising model,'' {\em Physical review letters}, vol.~95, no.~24,
  p.~245701, 2005.

\bibitem{dziarmaga2022coherent}
J.~Dziarmaga, M.~M. Rams, and W.~H. Zurek, ``Coherent many-body oscillations
  induced by a superposition of broken symmetry states in the wake of a quantum
  phase transition,'' {\em Physical Review Letters}, vol.~129, no.~26,
  p.~260407, 2022.

\bibitem{sorella2001generalized}
S.~Sorella, ``Generalized lanczos algorithm for variational quantum monte
  carlo,'' {\em Physical Review B}, vol.~64, no.~2, p.~024512, 2001.

\bibitem{torlai2018neural}
G.~Torlai, G.~Mazzola, J.~Carrasquilla, M.~Troyer, R.~Melko, and G.~Carleo,
  ``Neural-network quantum state tomography,'' {\em Nature Physics}, vol.~14,
  no.~5, pp.~447--450, 2018.

\bibitem{shi2019neural}
H.-Q. Shi, X.-Y. Sun, and D.-F. Zeng, ``Neural-network quantum state of
  transverse-field ising model,'' {\em Communications in Theoretical Physics},
  vol.~71, no.~11, p.~1379, 2019.

\bibitem{shi2022learning}
H.-Q. Shi and H.-Q. Zhang, ``Learning topological defects formation with neural
  networks in a quantum phase transition,'' {\em arXiv:2204.06769, DOI:
  10.1088/1572-9494/ad3227}, 2022.

\bibitem{cincio2007entropy}
L.~Cincio, J.~Dziarmaga, M.~M. Rams, and W.~H. Zurek, ``Entropy of entanglement
  and correlations induced by a quench: Dynamics of a quantum phase transition
  in the quantum ising model,'' {\em Physical Review A}, vol.~75, no.~5,
  p.~052321, 2007.

\end{thebibliography}

%\begin{thebibliography}{99}\footnotesize
%\vspace{2.5mm}
%\bibitem{carleo}

%\end{thebibliography}

\end{document}